\documentclass[11pt]{article}

\usepackage{lineno,hyperref,comment}
\modulolinenumbers[5]

\newcommand{\ignore}[1]{}
\usepackage{graphicx}
\usepackage{epsfig}
\usepackage{color}
\usepackage{amsmath}
\usepackage{amsthm}
\usepackage{epstopdf}
\usepackage{epsfig}

\newcommand{\Q}{{\bf Q}}

\newcommand{\magic}{\sqrt{{\cal P}}}

\definecolor{herecolor}{rgb}{0.1,0.6,0.1 }

\newcommand{\bartb}[1]{\textcolor{black}{#1}}
\newcommand{\bartwashere}[1]{\bigskip
\textcolor{herecolor}{\sc **** Bart has read till here ****}\bigskip\newline }

 \newcommand{\ab}{\allowbreak}
 
\newtheorem{definition}{Definition}
\newtheorem{example}{Example}

\newtheorem{theorem}{Theorem}
\newtheorem{lemma}{Lemma}

\newtheorem{property}{Property}

\def\squareforqed{\hbox{\rlap{$\sqcap$}$\sqcup$}}
\def\qedd{\ifmmode\squareforqed\else{\unskip\nobreak\hfil
\penalty50\hskip1em\null\nobreak\hfil\squareforqed
\parfillskip=0pt\finalhyphendemerits=0\endgraf}\fi}

\usepackage{listings}
\lstdefinestyle{tiny}{
  basicstyle=\sffamily\small, 
  numbers=none
  }

\newcommand{\out}[1]{}

\usepackage{pseudocode}  

\usepackage{makeidx}  
 \makeindex          

\renewcommand{\ni}{\noindent}

\date{}

\begin{document}

\title{A Formal Algebra for OLAP${}^1$}
\author{Bart Kuijpers${}^2$ and 
Alejandro Vaisman${}^3$}
\maketitle

\begin{abstract}
Online Analytical Processing (OLAP) comprises
 tools and algorithms that allow querying
multidimensional   databases. It is based 
on the multidimensional model, where   
data can be seen as a cube, where each cell
contains one or more  measures 
can be aggregated
along  dimensions. Despite the extensive corpus of work
in the field, a standard language for OLAP is still 
needed, since there is no well-defined, accepted semantics,
 for many 
of the usual OLAP operations. 
In this paper, we  address this problem, and present a set 
of operations for   manipulating a data cube. We  clearly 
define the semantics of these operations,   
and prove that they  can be composed, yielding a language  powerful enough 
to express complex OLAP  queries. We express these operations as a sequence of atomic 
transformations over  a fixed multidimensional matrix,
whose cells contain a sequence of measures. 
Each atomic transformation produces a new measure. When a sequence of transformations defines  an OLAP operation,  a flag  is produced  indicating which   
cells  must be considered as input for the next operation. In this way, an elegant  algebra is defined. Our main contribution, with respect to other similar efforts in 
the field is that, for the first time, a formal proof of the correctness of
the operations is given, thus providing a clear semantics for them. We believe 
the present work will serve as a basis to build more solid practical tools 
for data analysis.  
 \end{abstract}


\footnotetext[1]{Extended abstract. Full version to appear in Intelligent Data Analysis, 21(5), 2017.}
\footnotetext[2]{Databases and Theoretical Computer Science Research Group, Hasselt University and Transnational University of Limburg; email: bart.kuijpers@uhasselt.be}
\footnotetext[3]{Instituto Tecnol{\'o}gico de Buenos Aires, Buenos Aires, Argentina; email:  avaisman@itba.edu.ar. 
}.
 


\section{Introduction}\label{sec:intro}

Online Analytical Processing(OLAP)~\cite{Kim96} comprises
a set of tools and algorithms that allow efficiently querying
multidimensional (MD) databases  containing large amounts of data,
usually called Data Warehouses (DW). 
Conceptually, in the MD model,  
data can be seen as a {\em cube}, where each cell
contains one or more \textit{measures}
of interest, that quantify \textit{facts}. Measure values
can be aggregated
along \textit{dimensions}, which give context to facts. 
At the logical level, OLAP data are  typically organized as a set of
\textit{dimension and fact tables.} 
Current database technology allows  
alphanumerical warehouse data to be integrated 
for example,  with   geographical or social network data, for decision making.
In the era of so-called ``Big Data'', the kinds of data that could be handled
by data management tools, are likely to 
increase in the near future. Moreover, 
OLAP and Business Intelligence (BI) tools allow to capture, integrate, manage, and 
query, different kinds of  information.  
For  example,  alphanumerical  data coming from a local DW, spatial data (e.g., temperature) represented as rasterized images, and/or economical data published on the semantic web. 
 Ideally, a  BI user would just like to deal with what she knows well, namely the data cube, using only the classical OLAP operators, like \textit{Roll-up}, \textit{Drill-down}, \textit{Slice}, and \textit{Dice} (among other ones), regardless   the cube's  underlying data type. Data types should only be handled  at the logical and physical levels, not at the conceptual level. Building on this idea,  Ciferri et al.~\cite{CCG+13}  proposed a \textit{conceptual}, \textit{user-oriented} model,  independent of OLAP 
technologies.  In this model, the user only manipulates a data cube. Associated with the model, there is a query language providing high-level operations over the cube.  This language,  called Cube Algebra, was sketched informally in the mentioned work. Extensive examples on the use of Cube Algebra  presented in~\cite{VZ14}, suggest  that this idea can lead to a language  much more intuitive and simple than  
 MDX, the \textit{de facto} standard for OLAP. Nevertheless, these works do not give  
  any evidence of the correctness of the languages and operations proposed, other than examples at various degrees of comprehensiveness. In fact, surprisingly, and  in spite of the large corpus of work in the field, a formally-defined reference language for OLAP is still 
needed~\cite{RA07}. There is not even a well-defined, accepted semantics, for many of the usual OLAP operations.  We believe that, far for being just a problem of classical OLAP, this formalization is also needed in current ``Big Data'' scenarios, where there is a need to efficiently perform real-time OLAP operations~\cite{Dehne201531}, that, of course, must be well defined.

\paragraph*{Contributions}
In this paper we  (a)
introduce  a collection of operators that manipulate a data cube, and clearly define their  semantics; and
(b) prove, formally, that our operators can be composed, yielding a language  powerful enough to express 
complex  queries and cube navigation (``\textit{\`a la} OLAP'') paths.
 
We achieve the above representing the data cube as a fixed $d$-dimensional matrix, and a set of $k$ measures, and   expressing each OLAP operation as a sequence of atomic transformations. Each transformation produces a new measure, and, additionally, when a sequence forms an OLAP operation, a flag that indicates which are the cells that must be considered as input for the next operation. This formalism allows us to elegantly define an algebra as a collection of operations, and give a series of properties that show  their correctness. We provide the proofs in the full paper. We limit ourselves to the most usual operations, namely slice, dice, roll-up and drill-down, which constitute the core of all practical OLAP tools. We denote these the \textit{classical OLAP operations}. This allows us to focus on our main interest, which is, to prove the feasibility of the approach. Other not-so-usual operations are left for future work.

The main contribution of our work, with respect to other similar efforts in the field is that, for the first time, a formal proof to practical problems is given, so the present work will serve as a basis to build more solid  tools for data analysis. Existing work either lacks of formalism, or of applicability, and no work of any of these kinds give sound mathematical prove of its claims. In this extended abstract we present the main properties, and leave the proofs for the full paper. 


The remainder of the paper is organized as follows. 
In Section~\ref{sec:data-model}, we present our MD 
data model, on which we base the rest of our work. 
Section~\ref{sec:OLAP-TandO}  presents the atomic 
transformations that we use to build the OLAP operations. 
In Section~\ref{sec:classicalOLAP} we discuss the 
classical OLAP operations in terms of the transformations,
show how they can be composed to address complex queries. 
We conclude in 
Section~\ref{sec:Conclusion}.


\section{The OLAP Data Model}\label{sec:data-model}

In this section  we describe the {OLAP data model} we use in the sequel. 

\subsection{Multidimensional Matrix}\label{subsec:matrix}
 
We next give the definitions of  multidimensional matrix schema and  instance.
In the sequel, $d$, with $d\geq 1$, is a natural number representing the number of dimensions of a data cube.

\begin{definition}[Matrix Schema]\rm \label{def:matrix-schema}
A \emph{$d$-dimensional matrix schema} is a sequence $(D_1, \ab D_2,\ab  ...,\ab  D_d)$ of $d$ dimension names.
\qedd
\end{definition}

\bartb{Dimension names can be considered to be strings.}
As illustrated in the following example, the convention will be that dimension names  start with a capital letter.

\begin{example}\rm  \label{ex:matrix-schema}
The running example   we   use in this paper, deals with sales information of certain products, at certain locations, at certain moments in time. 
For this purpose, we will define a  $3$-dimensional matrix schema $(D_1, D_2, D_3)=(Product, \ Location,\  Time)$. 
\qedd
\end{example}

\begin{definition}[Matrix Instance]\rm  \label{def:matrix-instance}
A \emph{$d$-dimensional matrix instance} (\emph{matrix}, for short) over the $d$-dimensional matrix schema $(D_1, D_2, ..., D_d)$ 
is the product 
$dom(D_1)\times dom(D_2)\times \cdots \times dom(D_d),$  $i=1,2,..., d$,  where $dom(D_i)$  is   a non-empty, finite, ordered set, called  the \emph{domain}, that is  associated with the dimension name $D_i$. 
 For all   $i=1,2,..., d$,  we denote by $<$, the order that we assume on the elements of $dom(D_i)$.
 For $a_1\in dom(D_1), \ab a_2\in  dom(D_2), \ab ..., \ab a_d\in  dom(D_d)$, we call the tuple $(a_1, a_2, ..., a_d)$, a \emph{cell} of the matrix.
\qedd
\end{definition}

The cells of a matrix serve as placeholders for the measures that are contained in the data cube (see Definition~\ref{def:data-cube-instance} below).
Note that, as it is common practice in OLAP, we assumed an order $<$  on the domain. 
 The role of the order is  further  discussed in 
 Section~\ref{subsec:order}.
 
As a notational convention,   elements of the domains  $dom(D_i)$ start with a lower case letter, as it is shown in the following example.

\begin{example}\rm  \label{ex:matrix-instance}
For the $3$-dimensional matrix schema $(D_1, D_2, D_3)\ab =\ab (Product, \ \ab Location, \ \ab Time)$ of Example~\ref{ex:matrix-schema}, 
the non-empty sets $dom(D_1)=\{lego,\ab\ brio ,\ab\  apples,\ab \ oranges\}$, 
$dom(D_2)=\{antwerp, \ brussels, \ paris, \ marseille\}$, and $dom(D_3)=\{\mbox{1/1/2014}, \ab\ ..., \ab\ \mbox{31/1/2014}\}$ produce 
the matrix instance $dom(D_1)\times \ab dom(D_2)\times \ab  dom(D_3).$
 The cells of the matrix will contain the sales for each combination of values in the domain.
In $dom(D_2)$, we have, for instance, the order $antwerp < brussels < paris < marseille.$
Over the dimension $Time$, we  have the usual temporal order.
\qedd
\end{example}

\subsection{Level Instance, Hierarchy Instance and Dimension Graph}\label{subsec:instances}

We now define the notions of dimension schema  and   instance. 
 
\begin{definition}[Dimension Schema,  Hierarchy and Level]\rm  \label{def:dimension-schema}
Let $D$ be a name for  a dimension. A \emph{dimension \ab schema $\sigma(D)$ for $D$} is a lattice,  
with a unique top-node, called $All$ (which has only incoming edges) 
and a unique bottom-node, called $Bottom$ (which has only outgoing edges), 
such that all maximal-length paths in the graph go from $Bottom$ to $All$.
Any path from $Bottom$ to $All$ in a dimension schema $\sigma(D)$ is called a \emph{hierarchy} of $\sigma(D)$. 
Each node in a hierarchy (i.e., in a dimension schema) is called a \emph{level} (of $\sigma(D)$).
\qedd
\end{definition}
 As a  convention, level names start  with a capital letter. Note that the $Bottom$ node is often renamed, depending on the application.  

\begin{figure}[t]
\centering
   \centerline{\includegraphics[scale=0.5]{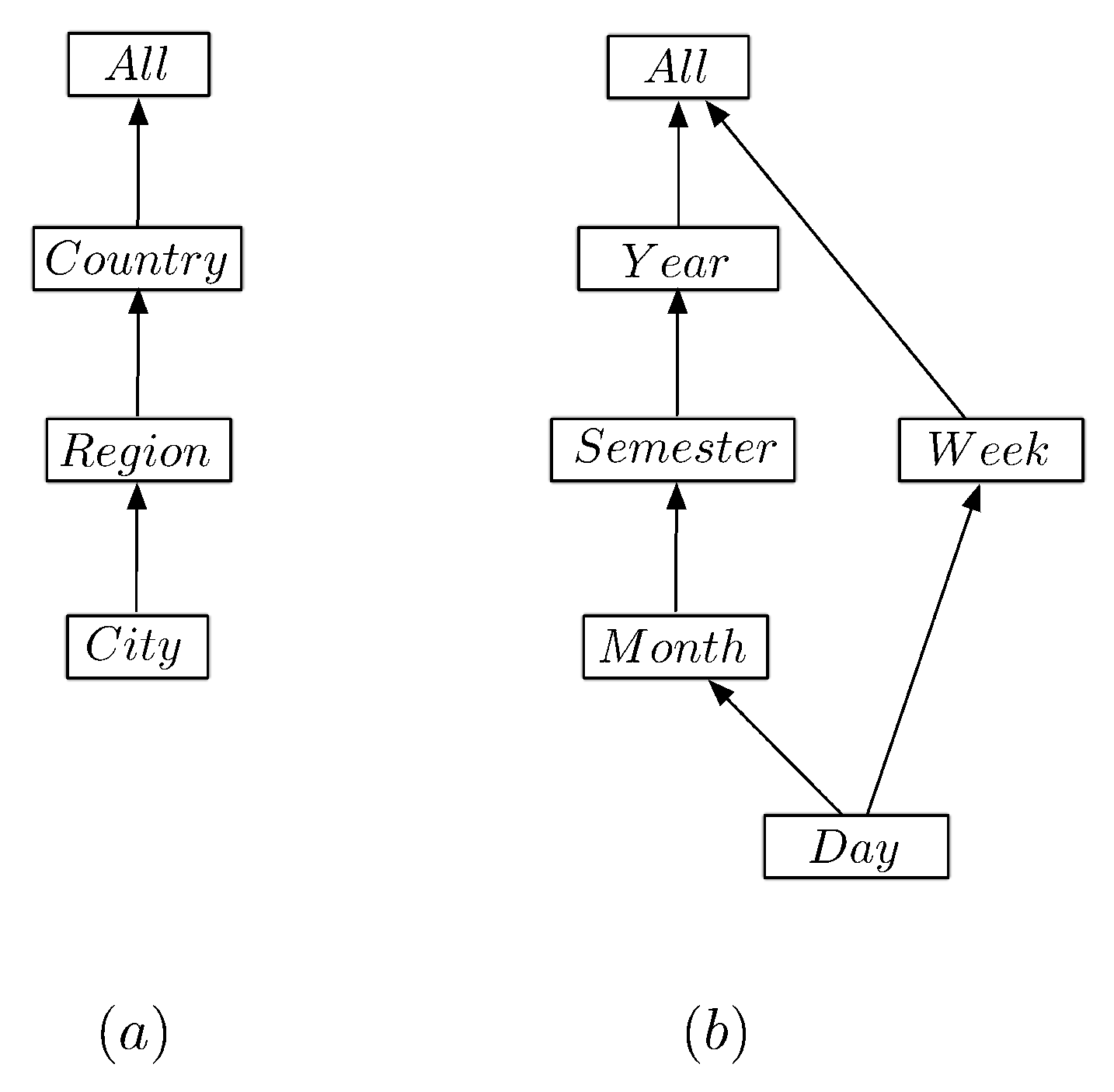}}
\caption{Dimension schemas for the dimensions $Location$, in ($a$), and $Time$ , in ($b$).}\label{fig:dimension-schema}
\end{figure}

\begin{example}\rm  \label{ex:dimension-schema}
Fig.~\ref{fig:dimension-schema} gives examples of dimension schemas $\sigma(Location)$ and $\sigma(Time)$ 
for the dimensions $Location$ and $Time$ in  Example~\ref{ex:matrix-schema}.
For the dimension $Location$, we have $Bottom=City$, and there is only one hierarchy,  denoted
$City\rightarrow Region \rightarrow Country\rightarrow All.$ The node $Region$ is an example of a level in this hierarchy.
For the dimension $Time$, we have $Bottom=Day$, and  two hierarchies, namely $Day\rightarrow Month \rightarrow Semester \rightarrow Year\rightarrow  All$
and $Day\rightarrow Week \rightarrow  All$.
\qedd
\end{example}

\begin{definition}[Level Instance, Hierarchy Instance, Dimension Graph]\rm \label{def:instanceGraph}
Let $D$ be a dimension with schema $\sigma(D)$, 
and let $\ell$ be a level  of $\sigma(D)$. A \emph{level instance of $\ell$ } is a non-empty, finite 
set $dom(D.\ell)$. If $\ell=All$, then $dom(D.All)$ is the singleton $\{all\}$. If $\ell=Bottom$, 
then $dom(D.Bottom)$ is the the domain of the dimension $D$, 
that is,  
$dom(D)$ (as in Definition~\ref{def:matrix-instance}).  

A \emph{dimension graph (or instance)} $I(\sigma(D))$ \bartb{over} the dimension schema $\sigma(D)$ is a directed acyclic graph with node set
$\bigcup_{\ell} dom(D.\ell),$ 
where the union is taken over all levels in $\sigma(D)$. The edge set of this directed acyclic graph
is defined as follows. Let $\ell$ and $\ell'$ be two levels of $\sigma(D)$, and let $a\in dom(D.\ell)$ and  $a'\in dom(D.\ell')$.
 Then, only if there is a directed edge from $\ell$ to 
$\ell'$ in $\sigma(D)$, there can be a directed edge in $I(\sigma(D))$ from $a$ to $a'$.

If $H$ is a hierarchy in $\sigma(D)$, then the \emph{hierarchy  instance} (relative to the dimension instance $I(\sigma(D))$)
is the subgraph   of $I(\sigma(D))$ with nodes from $dom(D.\ell)$, for $\ell$ appearing in $H$. This subgraph is denoted 
 $I_H(\sigma(D))$.
\qedd
\end{definition}

As notational convention,  the names of objects in a set $dom(D.\ell)$ start with  a  lower case character. 
We remark that a hierarchy instance $I_H(\sigma(D))$ is always a (directed) tree. 
 Also,  if $a$ and $b$ are two nodes in a hierarchy instance $I_H(\sigma(D))$, 
such that $(a,b)$ is in the transitive closure of the edge relation of $I_H(\sigma(D))$, 
 we will say that $a$ \emph{rolls-up} to $b$ and we  denote this by $\rho_H(a,b)$ (or $\rho(a,b)$  if $H$ is clear from the context).  
 
\begin{figure}[t]
\centering
   \centerline{\includegraphics[scale=0.5]{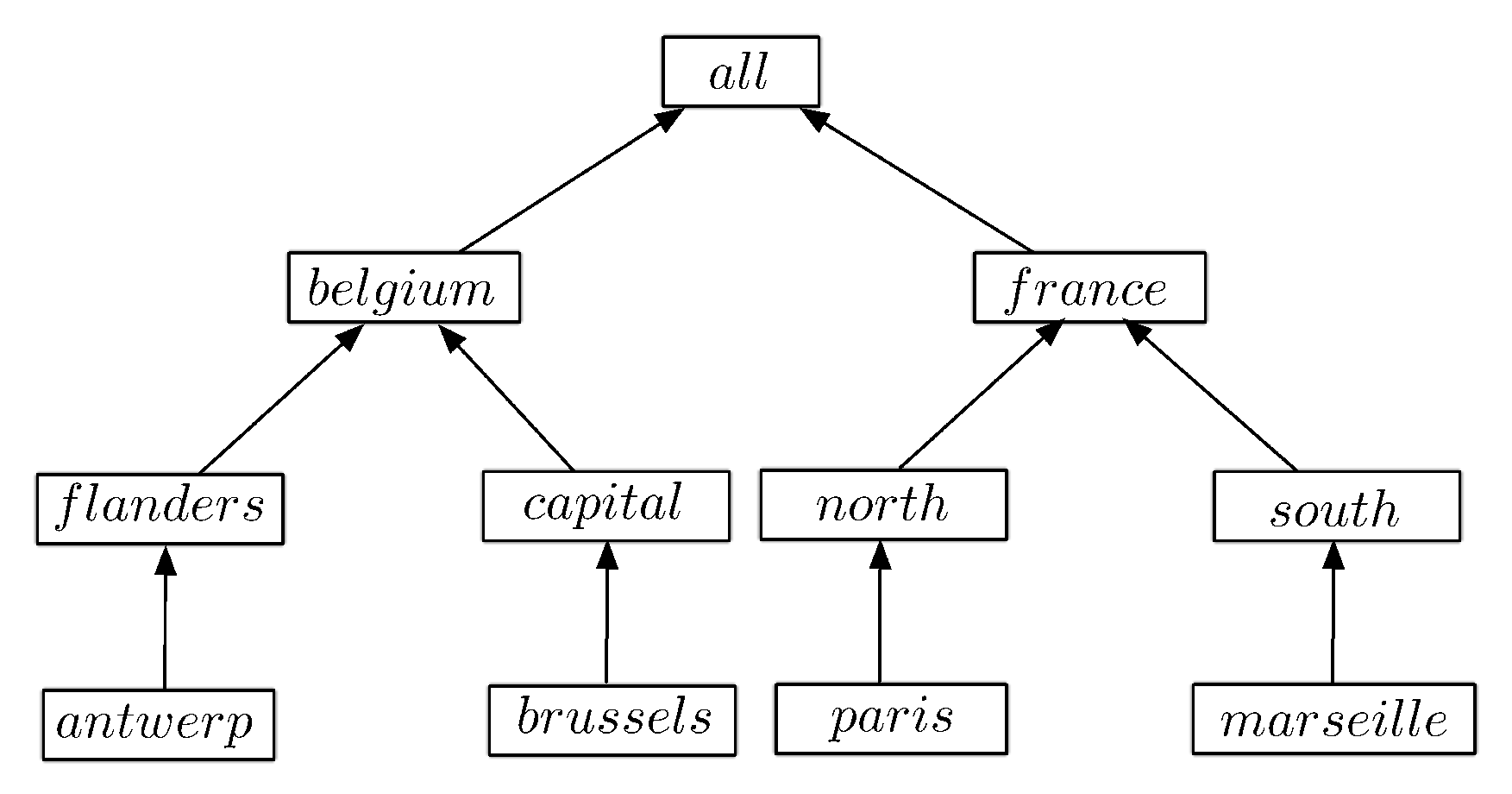}}
\caption{An example of a dimension graph \bartb{(or instance)} $I(\sigma(Location))$.}\label{fig:dimension-instance}
\end{figure}

 \begin{example}\rm  \label{ex:instanceGraph}
Consider the  $Location$ dimension, whose schema $\sigma(Location)$  is given in  Fig.~\ref{fig:dimension-schema} ($a$).
 From Example~\ref{ex:matrix-instance}, we have  
$dom(Location)=\{antwerp,$ $\ brussels,$ $ \ paris,$ $\ marseille\}$, which is  $dom(Location.Bottom)$, or $dom(Location.City)$.

 An example of a dimension instance $I(\sigma(Location))$ is depicted in Fig.~\ref{fig:dimension-instance}. 
 This example expresses, for instance, that the city $brussels$ is located in the region $capital$ which 
 is part of the country $belgium$, meaning that   $brussels$ rolls-up to $capital$ and to $belgium$, that is, $\rho(brussels, \ captial)$ and 
 $\rho(brussels, \ belgium)$.
\qedd
\end{example}

In a dimension graph, we must guarantee that rolling-up   through different paths   gives the same results. This is formalized by the concept of  ``sound'' dimension graph.
  
  \begin{definition}[Sound Dimension Graph]\rm \label{def:sound-Graph}
Let  $I(\sigma(D))$ be a dimension graph (as in Definition~\ref{def:instanceGraph}). We call this dimension graph \emph{sound}, if for any level $\ell$ 
 in $\sigma(D)$ and any two hierarchies $H_1$ and $H_2$ that reach $\ell$ from the $Bottom$ level and any $a\in dom (D)$ and $b_1,b_2\in dom(D.\ell)$, we have 
 that $\rho_{H_1}(a,b_1)$ and $\rho_{H_2}(a,b_2)$ imply that $b_1=b_2$.
  \qedd
\end{definition}
  
In this paper, we assume that dimension graphs are always sound.

 \subsection{Multidimensional Data Cube}\label{subsec:data-cube}
 
Essentially, a data cube  is a matrix in which the cells are filled with measures that are taken from some 
\emph{value domain} $\Gamma$. For many applications, $\Gamma$ will be the set of real or rational numbers, although some other 
ones may  include, e.g.,  spatial regions or geometric objects.  

\begin{definition}[Data Cube Schema]\rm \label{def:data-cube-schema}
A \emph{$d$-dimensional data cube schema} consists of 
 (a)  a $d$-dimensional matrix schema $(D_1, D_2, ..., D_d)$; and 
(b)  a hierarchy schema $\sigma(D_i)$ for each  dimension $D_i$, with $i=1,2,...,d$.  
\qedd
\end{definition}

\begin{definition}[Data Cube Instance]\rm \label{def:data-cube-instance}
Let $\Gamma$ be a non-empty set of ``values''.
A \emph{$d$-di\-men\-sio\-nal, $k$-ary data cube instance} (or \emph{data cube}, for short) $\cal D$
over the $d$-dimensional matrix schema $(D_1, D_2, ..., D_d)$ and hierarchy schemas $\sigma(D_i)$ for $D_i$, for $i=1,2,...,d$, with values from $\Gamma$,
  consists of 
 (a)  a $d$-dimensional matrix instance over the matrix schema $(D_1, D_2, ..., D_d)$,  $M({\cal D})$; 
 (b) for each $i=1,2,...,d$, a \textit{sound} dimension  graph $I(\sigma(D_i))$ over $\sigma(D_i)$;  
 (c) $k$ \emph{measures} $\mu_1,\mu_2,..., \mu_k$, which are functions from $dom(D_1)\times dom(D_2)\times \cdots \times dom(D_d)$
to the  value domain $\Gamma$; and 
 (d)  a \emph{flag} $\varphi$ , which is  a function from $dom(D_1) \times \cdots \times dom(D_d)$
to the set $\{0,1\}$.
\qedd
\end{definition}

\bartb{In the remainder of this paper  we assume that $\Gamma=\Q$, the set of the rational numbers. For most applications, 
this suffices.} Also, as a notational convention, we use calligraphic characters, like $\cal D$, to represent data cube instances.

The flag $\varphi$ can be considered as a $(k+1)$-st Boolean measure. 
The role of $\varphi$ is to indicate which of the matrix cells are currently ``active''. The active cells have a 
flag value $1$ and the others have a flag value $0$. When
we operate over a data cube, flags are used to indicate the input or output parts of the matrix of the cube.  
Typically, in the beginning of the operations, 
all cells have a flag value of $1$. 
The role of flags will become more clear in the next sections, when we discuss OLAP transformations and operations.

\subsection{Ordered Domains and the Representation of Higher-level Objects}\label{subsec:order}

When performing OLAP transformations and operations, 
we may need to store aggregate information about certain measures
up to some level above the $Bottom$ one. We do not want to 
use extra space for this in the data cube. Instead, we use
the available cells of 
the original data cube to store this  information. For this, we make use of the order
assumed in  Definition~\ref{def:matrix-instance}, for  the representation of  high-level objects by 
$Bottom$-level objects.
 

\begin{definition}\rm\label{def:represents}
Let $D\in \{D_1, D_2, ..., D_d\}$ be an arbitrary dimension with domain $dom(D)=dom(D.Bottom)$. 
Let $\ell$ be a level of $\sigma(D)$.   
An element $b\in dom(D.\ell)$ is \emph{represented} by the smallest 
element $a\in dom(D)$ (according to $<$) for which $\rho(a,b)$ holds. We denote this as \bartb{$rep(b)=a$}, and say that \bartb{\emph{$a$ represents $b$}}.\qedd 
\end{definition}

 \begin{example}\rm \label{ex:order}
Continuing with the previous examples, we consider the dimension $Location$
with $dom(Location)=\{antwerp, \ \ab brussels, \ \ab paris, \ \ab marseille\}$ (i.e., $dom(Location.City)$.
On this set, we \textit{assume} the order $antwerp < brussels < paris < marseille$. 
For this dimension, we have the hierarchy and   
 the dimension instance, given in Figs.~\ref{fig:dimension-schema} and ~\ref{fig:dimension-instance}, respectively.  
 At the $Bottom=City$ level, cities represent themselves. At higher levels, regions and countries are represented by their ``first'' city in 
$dom(Location)$ (according to $<$). 
Thus, $flanders$  and $belgium$ are 
represented by $antwerp$,  $france$ is represented by $paris$, and $south$ is represented by $marseille$.
At the level $All$,   $antwerp$   represents $all.$
\qedd
\end{example}


 Note that   the  $Bottom$-level  representatives  of higher-level objects, will be flagged $1$, and other  
 cells flagged $0$.  
 Also,  in our example, if we aggregate information at   level $Region$, 
 with $dom(Location.Region)=\ab \{flanders, \ \ab capital, \ \ab north, \ \ab south\}$, then all cities in $dom(Location)$ 
 become flagged.  Thus, it would not be clear if the cube contains information at the level $City$ or at the level $Region$. 
 \bartb{To solve this, we could keep a log of the OLAP operations that are performed, making the level of aggregation clear.}
 The following property shows how the order on the $Bottom$ level induces and order on higher levels. 

\begin{property}\rm\label{prop:order-higher-level}
Let $D\in \{D_1, D_2, ..., D_d\}$ be a (sound) dimension of a data cube $\cal D$ and let $\ell$ be a level in the dimension schema $\sigma(D)$.
The order $<$ on $dom(D)$ induces an order  on $dom(D.\ell)$ as follows. If $b_1,b_2\in dom(D.\ell)$, then 
$b_1<b_2$ if and only if $rep(b_1)<rep(b_2)$.\qedd
\end{property}



\section{OLAP Transformations and Operations}\label{sec:OLAP-TandO}
A typical OLAP user manipulates a data cube by means of well-known operations. 
For instance, using our running example,  the query ``Total sales by region, for regions in Belgium or France'', is actually expressed as a sequence of operations, whose semantics should be clearly defined, and which can be applied in different order.  
For example, we can first apply a  \textit{Roll-Up} (i.e., an aggregation) to the \textit{Country} level, and once at that level apply a \textit{Dice} operation, which keeps the cube cells corresponding to Belgium or France. Finally, a \textit{Drill-Down} can be applied to disaggregate  the sales down to the  level \textit{Region}, returning the desired result. 
In what follows, we characterize OLAP operations as the result of sequences of \bartb{``atomic''} OLAP transformations, which 
are measure-creating updates to a data cube. 
 
\subsection{Introduction to OLAP Transformations and Operations}
\label{subsec:informal-transfomation-operation}
 An \emph{atomic OLAP transformation} acts on a  data cube instance, by adding a    
 measure to the existing  
 data cube measures. OLAP operations like the ones informally introduced above 
 are defined, in our approach,  as a sequence of transformations. The 
 process  of OLAP transformations starts from a given \emph{input data cube} 
 \bartb{${\cal D}_{in}$}. 
 We assume that this original data cube  
 has $k$ given measures $\mu_1,\mu_2,..., \mu_k$ (as
 in Definition~\ref{def:data-cube-instance}).
 These $k$ measures have a special status in the sense that they are ``protected'' and can never be altered (see Section~\ref{subsec:OLAP-operation}).
  \bartb{However, there is one exception to this protection. These original measures can be ``destroyed'' in some cells (see further on), for instance, as the result of slice or dice operations, which are destructive by nature. Operations of these types destroy the content of some matrix cells and remove even the protected measures in it.} 
 
 
 Typically, the input-flag $\varphi$ of the original data cube ${\cal D}_{in}$ is set to $1$ in every cell and signals that every cell of $M({\cal D}_{in})$ is part of the input cube. 
 
Atomic OLAP transformations can be applied to data cubes. They add (or create) new measures to the sequence of existing measures by adding new measure values in each cell of the data cube's matrix. At any moment in this process, we may assume that the data cube $\cal D$ has 
 $k+l$ measures
 $\mu_1,\mu_2,..., \mu_k; \tau_1,...,\tau_l$, where the first  $k$ are the original measures of ${\cal D}_{in}$, 
 and   the last $l$ (with $l\geq 0$) ones
   have been created subsequently by $l$ OLAP  transformations (where $\tau_1,..., \tau_l$ is the empty sequence of $\tau$'s, for $l=0$). 
  The next OLAP transformation adds a new measure $\tau_{l+1}$ to the matrix cells.

We have said that we use OLAP transformations to compute 
 OLAP operations.  We indicate that the computation of an OLAP operation $O$ is finished by creating an $m$-ary output flag $\varphi^{(m)}_{O}$.  This output flag is a Boolean measure, that is created via  atomic OLAP transformations.
 It indicates which of the cells of $M({\cal D})$ should be considered as belonging to the output of $O$.
 It is $m$-ary in the sense that it keeps the last $m$ created measures $\tau_{l-m+1}, \tau_{l-m+2}, ..., \tau_{l}$ and 
 ``trashes''  the rest. It also removes the previous flag, which it replaces. 
  The initial measures $\mu_1,\mu_2,..., \mu_k$ of the input data cube ${\cal D}_{in}$ are never 
  removed  (unless they are ``destroyed'' in some cells). 
 They  remain in the cube 
 throughout the process of applying one OLAP operation after another to ${\cal D}_{in}$, and can be used  at any stage. 
  Summarizing,  after an OLAP operation of output arity $m$ is completed on some cube $\cal D$, 
 the measures in the cells of the output data cube ${\cal D}'=O({\cal D})$ are of the form 
 $ \underline{\mu_1,\mu_2,..., \mu_k} ; \tau_{l-m+1}, \tau_{l-m+2}, ..., \tau_{l}; \varphi^{(m)}_{O}.$
 Here,  the underlining indicates the protected status of these measures.
 After each OLAP operation, \bartb{we do a ``cleaning'' by renaming} the unprotected measures with the symbols $\tau_{1}, \tau_{2}, ..., \tau_{m}$ and the output measures become 
 $ \underline{\mu_1,\mu_2,..., \mu_k} ; \tau_{1}, \tau_{2}, ..., \tau_{m}; \varphi^{(m)}_{O}.$
 The next OLAP operation $O'$ can then act on   ${\cal D}'$ and use in its computation all the measures above. 
 We remark that the dimensions, the hierarchy schemas and instances of $\cal D$ remain unaltered during the entire OLAP process. 
 
 We end this description with a remark on \emph{destructors}.
 A destructor, optionally, precedes the creation of an output flag.
 A destructor $\delta$ takes the value $1$ for some cells of the matrix of a data cube, and $0$ on other cells. When $\delta$ is invoked (and activated by the 
 output flag that follows it)
 on a data cube $\cal D$ with measures $ \underline{\mu_1,\mu_2,..., \mu_k} ; \tau_{1}, \tau_{2}, ..., \tau_{m}$ and flag $ \varphi^{(m)}_{O}$, 
 it empties all cells for which the value of the destructor $\delta$  is $0$ by removing all measures from them, even the protected ones, 
 thereby effectively ``destroying'' these cells.
 This is the only case where the protected measures are altered (see operations Slice or Dice, later). 
  The output of a destructive operation $O$ looks like 
 $\underline{\mu_1,\mu_2,..., \mu_k} ; \tau_{1}, \tau_{2}, ..., \tau_{l}; \delta;  \varphi^{(m)}_{O}, $ 
 in which the destructor precedes the output flag. 
 The effect of the presence of a destructor is the following.
 A cell such that $\delta=0$ is emptied, after which it contains no more measures  and  flag.
 For cells with $\delta=1$, the sequence of measures
 $ \underline{\mu_1,\mu_2,..., \mu_k} ; \ab \tau_{1}, \tau_{2}, ..., \tau_{l}; \ab \delta; \varphi^{(m)}_{O}; $
 is transformed to $ \underline{\mu_1,\mu_2,..., \mu_k} ; \ab \tau_{l-m+1}, \ab \tau_{l-m+2}, \ab ..., \ab \tau_{l}; \ab \varphi^{(m)}_{O}; $ 
 which is renamed as $ \underline{\mu_1,\mu_2,..., \mu_k} ; \tau_{1}, \tau_{2}, ..., \tau_{m}; \varphi; $ 
 before the next transformation takes place. This transformation 
 will  act, cell per cell, on the matrix of a cube, and it does nothing with emptied cells. 
 That is, no new measure can ever be added to a destroyed cell. 
 

 
The following definition  specifies how an  OLAP transformation acts on a data cube. We then address in detail each atomic OLAP transformation  appearing in this definition.
  
   \begin{definition}[OLAP Transformation]\rm \label{def:OLAP-transformation}
  
  Let $\cal D$ be a $d$-dimensional, $(k+l)$-ary data cube instance   
  with given (or protected) {measures} $\mu_1,\mu_2,..., \mu_k$, created measures 
  $\tau_1,...,\tau_l$ (with $l\geq 0$) and flag $\varphi$ over some value domain $\Gamma$. 
  An \emph{OLAP transformation} $T$, applied to $\cal D$, results in  the creation of a new measure
  $\tau_{l+1}$ in $\cal D$. Transformation $T$ adds   measure  $\tau_{l+1}$ \bartb{to non-empty cells of $M({\cal D})$};   
  $\tau_{l+1}$
  is produced from: 
   $\mu_1,\mu_2,..., \mu_k$ \bartb{(in non-empty cells)}; 
    $\varphi$ \bartb{(in non-empty cells)};
   $\tau_1,\tau_2,...,\tau_l$ \bartb{(in non-empty cells)} and
    the hierarchy schemas and instances of $\cal D$;  
and belongs to one of the following classes: (a) Arithmetic transformations (Definition~\ref{def:trans-arith});
 (b) Boolean transformations (Definition~\ref{def:trans-boolean}); (c)  Selectors  (Definition~\ref{def:trans-selector});
 (d) Counting, sum, min-max (Definitions~\ref{def:trans-counting}, \ref{def:trans-min-max-revisited}); (e) 
 Grouping (Definition~\ref{def:trans-grouping}). 
 
An OLAP transformation can also result in the creation of a measure that is an output flag $\varphi^{(m)}$ of arity $m$. 
This should be a measure with a Boolean value.  To indicate that it is a flag of arity $m$, we use
the reserved symbol   $\varphi^{(m)}$ instead of $\tau_{l+1}$. 
 An output flag  $\varphi^{(m)}$ may (optionally) be preceded by a destructor $\delta$. 
This should be a measure with a Boolean value (to indicate which cells are destroyed). We use
the reserved symbol   $\delta$ instead of $\tau_{l+1}$.
\qedd
\end{definition}

\subsection{OLAP Operations and their Composition}\label{subsec:OLAP-operation}
Before we give the definition of an OLAP operation, we describe the \emph{input} to the OLAP process (this process may involve multiple OLAP operations). Such input is a $d$-dimensional, $k$-ary data cube instance 
${\cal D}_{in}$,  with measures $\mu_1,\mu_2,..., \mu_k$ and flag $\varphi$. 
 These measures are \emph{protected} in the sense that they remain the first $k$  
 measures throughout 
  the entire OLAP process and are never altered or removed unless they are destroyed in some cells. The cube ${\cal D}_{in}$ has also a Boolean flag $\varphi$, which typically has value $1$ in all cells of $M({\cal D}_{in})$. 
  Thus, the measures of the input cube ${\cal D}_{in}$ 
  are denoted 
  $\underline{\mu_1,\mu_2,..., \mu_k}; \varphi.$

 After applying  a sequence of  OLAP operations to ${\cal D}_{in}$, we obtain a data cube ${\cal D}$. 

   \begin{definition}[OLAP Operation]\rm 
  Let \bartb{$\cal D$} be a $d$-dimensional, $(k+l)$-ary \emph{input} data cube instance   
  with given {measures} $\mu_1,\mu_2,..., \mu_k$, computed measures $\tau_1,..., \tau_l$ and flag $\varphi$. 
  The data cube   $\cal D$ acts as  the input of an
   \textit{OLAP operation} $O$ (of arity $m$), which
  consists of a sequence of $n$ consecutive OLAP transformations that create the additional measures  
  $\tau_{l+1},..., \tau_{l+n}$, followed by the creation of an $m$-ary flag  $\varphi^{(m)}_{O}$ \bartb{(possibly preceded by a destructor $\delta$)}.
  As the result of the creation of $\varphi^{(m)}_{O}$, the measures in the cells of the data cube
are changed from $\underline{\mu_1,\mu_2,..., \mu_k} ; \tau_{1},..., \tau_l;\varphi;\tau_{l+1},..., \tau_{l+n}$ to
  $\underline{\mu_1,\mu_2,..., \mu_k} ; \tau_{l+n-m+1},..., \tau_{l+n}; \varphi^{(m)}_{O},$ 
  which become 
  $\underline{\mu_1,\mu_2,..., \mu_k} ; \tau_{1},..., \tau_{m}; \varphi,$  after renaming.  The output cube ${\cal D}'=O({\cal D})$ has the same dimensions, hierarchy schemas and instances as ${\cal D}$,
  and measures $ \underline{\mu_1,\mu_2,..., \mu_k} ; $ $\tau_{1},..., \tau_{m}; \varphi.$
 In the case where $\varphi^{(m)}_{O}$ is preceded by a destructor $\delta$, the same procedure is followed, except for the cells of $M({\cal D})$
  for which $\delta$ takes the value $0$. These cells of $M({\cal D})$ are emptied, contain no  measures, and become inaccessible for future 
  transformations. 
  \qedd
    \end{definition}

 
\subsection{Atomic OLAP Transformations}\label{subsec:atomicOLAP}

 We now address the five classes of atomic OLAP transformations of Definition~\ref{def:OLAP-transformation}.  We   use the following notational convention. 
 For a measure $\alpha$, we write  $\alpha(x_1,\ab x_2,  ..., \ab x_d) $ to indicate the value of  $\alpha$ in the cell $(x_1,\ab x_2,  ..., \ab x_d) \in dom(D_1)\times dom (D_2)\times \cdots \times dom(D_d).$  We remark that $\alpha(x_1,\ab x_2,  ..., \ab x_d) $ does not exist for empty cells and it is 
 thus not considered in computations. Also, we assume that there are \textit{protected} 
 measures $\mu_1,\mu_2,..., \mu_k$, and \textit{computed} measures $\tau_1,..., \tau_l$ in the non-empty cells, and call $\tau_{l+1}$ the next computed measure.

\subsubsection{Arithmetic Transformations}

\begin{definition}[Arithmetic Transformations]\rm\label{def:trans-arith}
The following creations of a new measure $\tau_{l+1}$ are \emph{arithmetic transformations}:
\begin{enumerate}
\item ({\bf \bartb{Rational} constant}) $\tau_{l+1}=\alpha$, with $\alpha\in\Q$, a rational number. 
\item ({\bf Sum}) $\tau_{l+1}=\alpha+\beta$, with $\alpha, \beta\in \{\mu_1,\mu_2,..., \mu_k, \tau_1,\ab \tau_2,\ab ...,\ab \tau_{l}\}$. 

\item ({\bf Product}) $\tau_{l+1}=\alpha\cdot\beta$, with $\alpha, \beta\in \{\mu_1,\mu_2,..., \mu_k, \tau_1,\ab \tau_2,\ab ...,\ab \tau_{l}\}$. 

\item ({\bf Quotient}) $\tau_{l+1}=\alpha/\beta$, with $\alpha, \beta\in \{\mu_1,\mu_2,..., \mu_k, \tau_1,\ab \tau_2,\ab ...,\ab \tau_{l}\}$. \bartb{Here, by convention,   
$a/0:=a$ for all $a\in \Q$.}
\qedd\end{enumerate}\end{definition}
  
\subsubsection{Boolean Transformations}

\begin{definition}[Boolean Transformations]\rm\label{def:trans-boolean}
The following creations of a new measure $\tau_{l+1}$ are \emph{Boolean transformations}:
\begin{enumerate}
\item ({\bf Equality test on measures}) $\tau_{l+1}=(\alpha=\beta)$, with $\alpha, \beta\in \{\mu_1, \ab \mu_2, \ab ..., \ab  \mu_k,  \ab \tau_1,\ab \tau_2,\ab ...,\ab \tau_{l}\}$. Here, the result of   $(\alpha=\beta)$
is a Boolean 1 or 0 (\bartb{cell per cell in the non-empty cells of the matrix}). 

\item ({\bf Comparison test on measures}) $\tau_{l+1}=(\alpha<\beta)$, with $\alpha, \beta\in \{\mu_1, \ab \mu_2, \ab ..., \ab \mu_k, \ab  \tau_1,\ab \tau_2,\ab ...,\ab \tau_{l}\}$. Here, the result of the comparison $(\alpha<\beta)$
is a Boolean 1 or 0 (\bartb{cell per cell in the non-empty cells of the matrix}).   

\item ({\bf Equality test on levels}) \bartb{For  a level $\ell$ in the dimension schema $\sigma(D_i)$ of dimension $D_i$,  and a constant object $c\in dom(D_i.\ell)$, 
$\tau_{l+1}(x_1,\ab x_2,  ..., \ab x_d) =(\ell =c)$ is an ``equality'' test. 
Here, the result of   $(\ell =c)$
is a Boolean 1 or 0 (\bartb{cell per cell in the non-empty cells of the matrix}) such that $\tau_{l+1}(x_1,\ab x_2,  ..., \ab x_d) $
is $1$ if and only if $x_i$ rolls-up to $c$ at level $\ell$, that is $\rho(x_i,c)$. }

\item ({\bf Comparison test on levels}) 
For a level $\ell$ in the dimension schema $\sigma(D_i)$ of dimension $D_i$,  and a constant   $c\in dom(D_i.\ell)$,  
$\tau_{l+1}(x_1,\ab x_2,  ..., \ab x_d) =(\ell <_{\ell}c)$ is a ``comparison'' test. 
 The result of  $(\ell <_{\ell}c)$
is a Boolean 1 or 0 (\bartb{cell per cell in the non-empty cells of the matrix}), such that $\tau_{l+1}(x_1,\ab x_2,  ..., \ab x_d) $
is $1$ if and only if $x_i$ rolls-up to an object $b$ at level $\ell$ for which $b<_{\ell} c$. The order $<_{\ell}$ can be any order that is defined on level $\ell$. 
 Transformation $\tau_{l+1}(x_1,\ab x_2,  ..., \ab x_d) =(c <_{\ell}\ell)$ is defined similarly.  
\qedd\end{enumerate}
\end{definition}


\begin{example}\rm\label{ex:trans-boolean-1}
We illustrate the use of Boolean transformations by means of a sequence of transformations that 
implement a ``dice'' (see Section~\ref{subsec:Dice} for more details).   The query
$\mbox{\sf DICE}({\cal D}, sales>50)$ asks for the cells  in the matrix of $\cal D$ which contain sales that are higher than 50.
 This query can be implemented by the  following sequence of transformations:
\begin{itemize}
\item $\tau_{1}= 49.99$ (rational constant);
\item $\tau_{2}= (\tau_{1}<sales)$ (comparison test on measures);
\item $\tau_{3}= \mu_1\cdot \tau_{2}$ (product);
\item $\delta=\tau_2$ (destructor); and
\item $\varphi^{(1)}=\tau_2$ (unary flag)
\end{itemize}

The measure $\tau_3$ contains the $sales$ values larger than  or equal to 50 
(and a 0 if the $sales$ are lower). The destructor $\delta$ destroys the cells that contain a O. Finally, the flag $\varphi^{(1)}$ selects all cells from the input as output cells (it will contain a 1 for all such cells that satisfy the condition), and concludes the $\mbox{\sf DICE}({\cal D}, sales>50)$ operation.
The output of this operation is $\underline{sales}; \tau_3; \varphi^{(1)},$ which is then renamed to $\underline{sales}; \tau_1; \varphi.$
\qedd\end{example}

\subsubsection{Selectors}
\begin{definition}[Selector Transformations]\rm\label{def:trans-selector}
The following creations of a new measure $\tau_{l+1}$ are \emph{selector transformations} (or \emph{selectors}), and their definition is  
cell per cell of $M({\cal D})$:
\begin{enumerate}
 \item  ({\bf Constant selector})  For a level $\ell$ in the dimension schema $\sigma(D_i)$  of a   dimension $D_i$, and  $c\in dom(D_i.\ell)$, 
 $\tau_{l+1}$ can be a \emph{constant-selector for $c$}, denoted $\sigma_{D_i.\ell=c}$, \bartb{ and it corresponds to the equality test on levels
 $\tau_{l+1}(x_1,\ab x_2,  ..., \ab x_d) =(\ell =c)$.}
 
 \ignore{which means 
that for all  $a\in dom(D_i)$ we have, for all $x_j\in dom(D_j)$ with $j\not= i$,   
$$\tau_{l+1}(x_1,..., x_{i_1},a, x_{i+1}, ..., x_d)= \left\{ \begin{array}{l l}
 1 &\mbox{ if }  \rho(a,b),\\
0 & \mbox{ otherwise. } \\
 
 \end{array}\right.$$ }

\item  ({\bf Level selector})  For a level $\ell$ in the dimension schema  $\sigma(D_i)$ 
of a   dimension $D_i$, $\tau_{l+1}$ can be a \emph{level-selector for $\ell$}, 
denoted by $\sigma_{D_i.\ell}$, which means 
that   we have, for all $x_j\in dom(D_j)$ with $j\not= i$,  
$$\tau_{l+1}(x_1,..., x_{i_1},a, x_{i+1},..., x_d) = \left\{ \begin{array}{l l}
 1 &\mbox{if}~a=rep(b)~\\
  &\mbox{for some}~b\in dom(D_i.\ell),\\
0 & \mbox{otherwise. } \\
  \end{array}\right.$$ 
\qedd\end{enumerate}
\end{definition}

The \textit{constant} selector in  Definition~\ref{def:trans-selector}, corresponds to the equality test on levels (see 3. in Definition~\ref{def:trans-boolean}).
Here, this transformation appears with a different functionality and we reserve a special notation for it, and we repeated it. Also, note  that the \textit{level} selector selects all representatives (at the $Bottom$ level) of objects at level $\ell$ of dimension $D_i$.



\begin{example}\rm\label{ex:trans-selector-2}
The query
$\mbox{\sf DICE}({\cal D}, Location.City = antwerp \ OR \ Location.City = brussels),$  asks for the sales in the cities of  $antwerp$ and $brussels$.   It can be implemented by the  following sequence of transformations, where $\tau_3$ can  take values $0$ or $1$, since the cities $antwerp$ and $brussels$ do not overlap:
 
  \begin{itemize}
\item $\tau_{1}= \sigma_{Location.City = antwerp}$ (constant selector);
\item $\tau_{2}= \sigma_{Location.City = brussels}$ (constant selector);
\item $\tau_{3}=  \tau_1+\tau_2$ (sum);
\item $\tau_{4}= \tau_3\cdot \mu_1$ (product);
\item \bartb{$\delta=\tau_3$ (destroys the cells outside $antwerp$ and $brussels$);}
\item $\varphi^{(1)}=\tau_3$ (unary flag creation).
\end{itemize}
\qedd\end{example}

\subsubsection{Count, Sum and Min-Max}
\begin{definition}[Counting, Sum, and Min-Max Transformations]\rm\label{def:trans-counting}
The   creations of a new measure $\tau_{l+1}$ defined next, are denoted  \bartb{\emph{counting, sum and min-max transformations}:}
\begin{enumerate}
  \item ({\bf Count-Distinct}) $\tau_{l+1}=\#_{\not=}(\alpha)$,  $\alpha \in \{\mu_1,\mu_2,..., \mu_k, \tau_1,\ab \tau_2,\ab ...,\ab \tau_{l}\}$ 
counts the number of distinct values of  measure $\alpha$ in the complete matrix $M({\cal D})$ of the data cube. 


\item  ({\bf $d$-dimensional sum})  $\tau_{l+1}=\sum_{(x_1,x_2, ..., x_d)\in M({\cal D})}
\alpha(x_1,x_2..., x_d),$ with $\alpha \in \{\mu_1,\mu_2,...,$ $\mu_k, \tau_1,\ab \tau_2,\ab ...,\ab \tau_{l}\}$,
gives the sum of the measure $\alpha$ over all \bartb{non-empty} matrix cells. We abbreviate this operation by  writing 
$\tau_{l+1}=\mbox{\sc SUM}_d(\alpha),$ and call this transformation the \emph{$d$-dimensional sum}.

\item ({\bf Min-Max})  $\tau_{l+1}=\min(\alpha)$, with $\alpha \in \{\mu_1,\mu_2,..., \mu_k, \tau_1,\ab \tau_2,\ab ...,\ab \tau_{l}\}$, gives the  
 smallest value of 
the measure $\alpha$ \bartb{in non-empty cells of }the matrix $M({\cal D})$. Similarly, $\tau_{l+1}=\max(\alpha)$, gives the largest value of 
the measure $\alpha$ in the matrix $M({\cal D})$. 
\qedd\end{enumerate}
\end{definition}

 It is important to remark that the above transformations create the \textit{same new measure value} for all cells of the matrix $M({\cal D})$.


\begin{example}\rm \label{ex:trans-counting-2}
Now, we look at the query $\mbox{``total sales in $antwerp$''.}$ 
The query can be computed as follows, given $\mu_1=sales$:
\begin{itemize}
 \item $\tau_{1}= \sigma_{Location.City = antwerp}$ (constant selector on $antwerp$);
  \item $\tau_{2}=\tau_1\cdot \mu_1$ (product that selects the sales in $antwerp$, puts a 0 in all  other ones);
  \item $\tau_{3}=\mbox{\sc SUM}_{3}(\tau_{2})$ (this is the total sales in $antwerp$ in every cell);
   \item $\tau_{4}= \tau_{3}\cdot \tau_1$  (this is the total sales in $antwerp$ in the cells of $antwerp$);
 \item $\varphi^{(1)} =\tau_1$ (this flag creation selects the cells of $antwerp$).
\end{itemize} 
The output measures are $\underline{sales}; \tau_4;\varphi^{(1)} $, which are renamed $\underline{sales}; \tau_1;\varphi $.
Thus, the value of the total of sales in $antwerp$ is now available in every cell corresponding to $antwerp$. For the cells outside $antwerp$ there is a $0$.
\bartb{We remark that this example can be modified with a destructor that effectively empties cells outside $antwerp$.}
\qedd
 \end{example}


\subsubsection{Grouping}

The most common   OLAP operations (e.g., roll-up, slice), require grouping data before aggregating them. For example, typically we will ask queries like ``total sales by city'', which requires grouping facts by city, and, for each group, sum 
all of its sales. Therefore, we need a transformation to express ``grouping''.  
To deal with grouping, we use the concept of ``prime labels'' for sets and products of sets. We will use these labels to identify elements in dimensions and in dimension levels.  
Before giving the definition of the grouping transformations, we elaborate on \bartb{prime labels and product of prime labels}. 
As we show, these prime labels work in the context of measures that take rational values  (as it is often the case, in practice).
The following definition specifies our infinite supply of prime labels.

\begin{definition}[Prime Labels]\rm \label{def:prime-labels}
Let $p_n$ denote the $n$-th prime number, for $n\geq 1$. 
 We define the sequence of \emph{prime labels}  as follows: $1, \ab \sqrt{2}, \ab \sqrt{3}, \ab \sqrt{5}, \ab \sqrt{7},\ab \sqrt{11}, \ab  ..., \ab \sqrt{p_n}, ....$
  We denote the set of all  prime labels by $\magic$.
\qedd
\end{definition}



\begin{definition}[Prime Labeling of Sets]\rm \label{def:prime-labelling-of-sets}
Let $A$, $A_1,A_2,..., A_n$ be (finite) sets.
A \emph{prime labeling} of the set $A$ is an injective function $w:A\rightarrow\magic$.
For $a\in A$, we call $w(a)$ the \emph{prime label} of $a$ (for the prime labeling $w$).

Let $I$ be a subset of $\{1,2,..., n\}$, which serves as an index set.
 A \bartb{\emph{prime product $I$-labeling}} of the Cartesian product $A_1\times A_2\times \cdots\times A_n$ 
consists of prime labelings $w_i$ of the sets $A_i$, for $i\in I$,  
that satisfy the condition that $w_i(A_i)\cap w_j(A_j)$ is empty for $i,j\in I $ and $i\not=j$.
For $(a_1,a_2,..., a_n)\in A_1\times A_2\times \cdots\times A_n$, we call 
$\prod_{i\in I}w_i(a_i)$ the \emph{prime product $I$-label}
of  $(a_1,a_2,..., a_n)$ (given the prime labelings $w_i$, for $i\in I$). When $I$ is a strict subset of $\{1,2,..., n\}$, we speak about a \bartb{\emph{partial prime product labeling}} and when 
$I=\{1,2,..., n\}$, we speak about a \bartb{\emph{full prime product labeling}}. 
\qedd
\end{definition}

If we view a Cartesian product $A_1\times A_2\times \cdots\times A_n$ as a finite matrix, whose cells contain rational-valued measures, we can use 
prime (product) labelings as follows in the aggregation process. Let us assume that the cells of $A_1\times A_2\times \cdots\times A_n$  contain rational values of a measure $\mu$ and let us denote 
the value of this measure in the cell $(a_1,a_2,..., a_n)$ by $\mu(a_1,a_2,..., a_n)$. If we have a full prime product labeling on $A_1\times A_2\times \cdots\times A_n$, then we can consider the sum over this Cartesian product
of the product of the prime product labels with the value of $\mu$:
$$\sum_{(a_1,a_2,..., a_n)\in A_1\times A_2\times \cdots \times A_n} \mu(a_1,a_2,..., a_n)\cdot w_1(a_1)\cdot w_2(a_2)\cdots w_n(a_n).\eqno{(\dagger_1)}$$

Since each cell of $A_1\times A_2\times \cdots\times A_n$ has a unique prime product label, and since these labels are rationally independent
(see Property~\ref{prop:prime-sum}), this sum  enables us to retrieve the values $\mu(a_1,a_2,..., a_n).$

If we have a partial prime product labeling on $A_1\times A_2\times \cdots\times A_n$, determined by an index set $I$, 
then, again,  we can consider the sum over this Cartesian product
of the product of the partial prime product labels with the value of $\mu$:
$$\sum_{(a_1,a_2,..., a_n)\in A_1\times A_2\times \cdots \times A_n} \mu(a_1,a_2,..., a_n)\cdot \prod_{i\in I}w_i(a_i).\eqno{(\dagger_2)}$$

Now, all cells in $A_1\times A_2\times \cdots\times A_n$ 
above a cell in the projection of $A_1\times A_2\times \cdots\times A_n$  \bartb{on its components with indices in $I$}, receive the same prime label.
This means that these cells are ``grouped'' together and the above sum allows us to retrieve the part of the sum that belongs to each group.
The following definition gives a name to the above sums.

\begin{definition}[Prime Sums]\rm \label{def:prime-sums}
\bartb{We call sums of type $(\dagger_1)$ \emph{full prime sums} and sums of type  $(\dagger_2)$ \emph{partial  prime sums (over $I$)}.}
 \qedd
\end{definition}

The following property can be derived from the well-known fact that the field extension $\Q(\sqrt{2}, \sqrt{3}, ..., \sqrt{p_{n}})=\{a_0+a_1\sqrt{2}+a_2 \sqrt{3}+\cdots +a_n\sqrt{p_{n}}\mid a_0,a_1, a_2,..., a_n\in \Q \}$ has degree 
$2^n$ over $\Q$ and corollaries of this property (see Chapter 8 in~\cite{escofier}). No square root of a prime number is a rational combination of square roots of other primes.

\begin{property}\rm\label{prop:prime-sum}
Let $n\geq 1$ and let $A_1\times A_2\times \cdots\times A_n$ be a Cartesian product of finite sets.
We assume that the cells $(a_1,a_2,..., a_n)$ of this set contain rational values $\mu(a_1,a_2,..., a_n)$ of a measure $\mu$.
Let $I$ be a subset of $\{1,2,..., n\}$ and let $w_i$ be prime labelings of the sets $A_i$, for $i\in I$, that form a prime product $I$-labeling.
Then, the prime sum $(\dagger_2)$ uniquely determines the values $\sum_{\times_{i\in I^c}A_i} \mu(a_1,a_2,..., a_n)$
for all cells of 
$ A_1\times A_2\times \cdots\times A_n$. \qedd 
\end{property}

 We remark that we use these prime (product) labels in a purely \emph{symbolic} way without actually  calculating the square root values in them. 
 We are now ready to define atomic OLAP operations that  allow us to implement grouping.
In what follows, we apply these prime labels to the case where the sets $A_i$ in $A_1\times A_2\times \cdots \times A_n$ are domains of dimensions (e.g., at the bottom level), 
or domains of dimensions at some level.

\begin{definition}[Grouping Transformations]\rm\label{def:trans-grouping}
The following creations of a new measure $\tau_{l+1}$ are \emph{grouping transformations}:
\begin{enumerate}
 \item ({\bf \bartb{Prime labels for groups in one dimension}}) Let $D_i$ be a dimension and $\ell$ 
 a level   in the dimension schema $\sigma(D_i)$  of a   dimension $D_i$.
Let $dom(D_i.\ell)=\{b_1, b_2, ..., b_m\}$  with induced order $b_1<b_2< \cdots< b_m$
(see Property~\ref{prop:order-higher-level}). 
 If the prime labels $w_1, w_2, ..., w_k$ have been used by previous transformations, then for \bartb{all $j$,  with $j\not=i$, and all $x_j\in dom (D_j)$}, 
 we have $\tau_{l+1}(x_1,...,x_{i-1},x_i,x_{i+1}, ..., x_d)=w_{k+l}$ if $\rho(x_i,b_l)$.
We denote this transformation by $\gamma_{D_i.\ell}(x_1,...,x_{i-1},x_i,x_{i+1}, ..., x_d)$ or $\gamma_{D_i.\ell}$, for short, 
and call the result of such a transformation a 
\emph{prime labeling}.

 \item  ({\bf Projection  of a prime sum})
 If the result of some previous transformation $\tau_m$ is a \bartb{(full or partial) prime sum } $\sum_{i=k}^{k+l} a_i\cdot w_i$ (over the complete matrix $M({\cal D})$)
 in which prime (product) labels $w_k, w_{k+1}, ..., w_{k+l}$ (computed in a previous transformation $\tau_n$) are used, then 
  $\tau_{l+1}$ is a new measure that ``projects'' on the appropriate component from the prime sum, that is, 
    $\tau_{l+1}(x_1,x_2 ..., x_d)=a_{k+l}$ if the prime (product) label $\tau_n(x_1,x_2 ..., x_d)= w_{k+l}$.
We denote this \bartb{projection transformation by $\tau_m\mid_{\tau_n}$}.
 \qedd\end{enumerate}
\end{definition}

\begin{example}\rm \label{ex:trans-grouping-3}
Consider the query $\mbox{``for each country, give the total
 number of}$ $\mbox{cities''.}$ 
This query can be implemented as follows (explained  below,  using the data  in Example~\ref{ex:instanceGraph}):

 \begin{itemize}
\item $\tau_{1}=\gamma_{Location.Country}$ (this gives each country a prime label);
 \item $\tau_{2}=\gamma_{Location.City}$  (this gives each city a \bartb{(fresh)} prime label);
  \item $\tau_{3}=\tau_{1}\cdot \tau_{2}$ (this gives each city a product of prime labels);
 \item $\tau_{4}= \mbox{\sc SUM}_{3}(\tau_{3})$;
  \item $\tau_{5}= \gamma_{Product.Bottom}$ (gives each product a different prime label);
  \item $\tau_{6}=\#_{\not=}(\tau_5) $ (counts the number of products); 
   \item $\tau_{7}= \gamma_{Time.Bottom}$ (gives each time moment a different prime label);
  \item $\tau_{8}=\#_{\not=}(\tau_7) $ (counts the number of moments in time); 
  \item $\tau_{9}=\tau_{6}\cdot \tau_{8}$ (is the number of products times the number of time moments);
 \item $\tau_{10}=\tau_{4}/\tau_{9}$ (normalization of the sum);
 \item $\tau_{11}=\tau_{10}\mid_{\tau_{2}}$; (projection over the prime labels of city);

 \item $\tau_{12}=\mbox{\sc SUM}_{3}(\tau_{11})$ (3-dimensional sum);

 \item $\tau_{13}=\tau_{12}/\tau_{9}$ (normalization of the sum);
 
  \item $\tau_{14}=\tau_{13}\mid_{\tau_{1}}$ (projection over the prime labels of country);
      \item $\varphi^{(1)} =\sigma_{Location.Bottom}$ (this flag creation selects all cells of the matrix).
\end{itemize}

 Transformation $\tau_{1}$ gives each country a next available prime label. Since no labels have been used yet, $belgium$ gets label $1$ and
 $france$ gets label $\sqrt{2}$. Transformation $\tau_{2}$ gives each city a next available prime label. Since $1$ and $\sqrt{2}$
  have been used, $antwerp$ gets label  $\sqrt{3}$, $brussels$ gets label  $\sqrt{5}$, $paris$ gets label  $\sqrt{7}$, and $marseille$ gets label  $\sqrt{11}$.

Transformation $\tau_{3}$ gives    $antwerp$ the value $\sqrt{3}$ (i.e.,
 $1.\sqrt{3}$, $brussels$  the value  $\sqrt{5}$($1.\sqrt{5}$), $paris$  the value   $\sqrt{14}$ ($\sqrt{2}.\sqrt{7}$), and $marseille$  the value  
  $\sqrt{22}$ ($\sqrt{2}.\sqrt{11}$). If there are 10 products and 100 time moments, then $\tau_4$ puts the value $10\cdot 100\cdot(\sqrt{3}+\sqrt{5}+\sqrt{14}+\sqrt{22})$ in each cell of the matrix $M({\cal D})$. 

 Transformations $\tau_6$ and $\tau_8$ count the number of products and the number of time moments (using fresh prime labels), and the product of these quantities is computed in
 $\tau_{9}$.  In $\tau_{10}$, 
$\tau_3$ is divided by this  product, putting $\sqrt{3}+\sqrt{5}+\sqrt{14}+\sqrt{22}$ in every cell. 

Transformation $\tau_{11}$ is a projection on the prime labels of $City$.
Since $\sqrt{3}$, $\sqrt{5}$, $\sqrt{7}$, and $\sqrt{11}$ are the prime  labels for the cities, and since  
 $\sqrt{3}+\sqrt{5}+\sqrt{14}+\sqrt{22}= 1\cdot \sqrt{3}+1\cdot \sqrt{5}+\sqrt{2}\cdot \sqrt{7}+\sqrt{2}\cdot \sqrt{11}$ , this will put  
 $1$ in the cells of $antwerp$ and $brussels$, and   $\sqrt{2}$ in the cells of $paris$ and $marseille$. 
 
 Next, $\tau_{12}$ puts  $10\cdot 100\cdot (2\cdot 1+2\cdot   \sqrt{2})$ in every cell of the cube and $\tau_{13}$ puts  
 $2\cdot 1+2\cdot   \sqrt{2}$ in every cell of the cube.
Finally, $\tau_{14}$ projects on the prime labels of countries, which are 1 and $\sqrt{2}$. 
  This puts a 2 in every cell of a Belgian city and a 2 in every cell in a French city. This is the result of the query, as the flag indicates,  that is returned in every cell. Now every cell 
  of a city in $belgium$ has the count of $2$ cities, as has every city in $france$.
 \qedd
 \end{example}

\subsubsection{Counting and Min-Max Revisited}

We can now extend the  transformations of Definition~\ref{def:trans-counting}, in a way that the counting,   minimum, and   maximum, are taken over cells which share a common prime product label.

\begin{definition}\rm\label{def:trans-min-max-revisited}

The following creations of a new measure $\tau_{l+1}$ are \bartb{generalizations of the \emph{counting and min-max} transformations}:
\begin{enumerate}
  \item ({\bf Count-Distinct}) 
  \bartb{If the result of some previous transformation $\tau_m$ is a prime (product) labeling of the cells of $M({\cal D})$, then
  $\tau_{l+1}(x_1,x_2 ..., x_d)=\#_{\not=}\mid_{\tau_m}(\alpha)$, with $\alpha \in \{\mu_1,\mu_2,..., \mu_k, \tau_1,\ab \tau_2,\ab ...,\ab \tau_{l}\}$
counts the number of different values of the measure $\alpha$ in cells of $M({\cal D})$ 
that have the same prime product label as  $\tau_m(x_1,x_2 ..., x_d)$.}

 \item ({\bf Min-Max})  
 \bartb{If the result of some previous transformation $\tau_m$ is a prime (product) labeling of the cells of $M({\cal D})$, then
$\tau_{l+1}(x_1,x_2 ..., x_d)=\min\mid_{\tau_m}(\alpha)$, with $\alpha \in \{\mu_1,\mu_2,..., \mu_k, \tau_1,\ab \tau_2,\ab ...,\ab \tau_{l}\}$, gives the the smallest value of
  the measure $\alpha$ in cells of the matrix $M({\cal D})$ 
that have the same prime product label as  $\tau_m(x_1,x_2 ..., x_d)$. And $\tau_{l+1}(x_1,x_2 ..., x_d)=\max\mid_{\tau_m}(\alpha)$ is defined similarly.}
\qedd\end{enumerate}
\end{definition}

 We remark that when there is only one prime label throughout $M({\cal D})$,   the above generalization
 of the counting and min-max transformations correspond to   Definition~\ref{def:trans-counting}. 

\section{The Classical OLAP Operations}\label{sec:classicalOLAP}

In this section, we prove that the classical OLAP operations can be expressed using the OLAP transformations from Section~\ref{sec:OLAP-TandO}. These classic operations can be combined to express complex 
analytical queries.  The classical OLAP operations are Dice,  Slice, Slice-and-Dice, Roll-Up  and 
 Drill-Down (see Section~\ref{subsec:RollUp}).
 We assume in the sequel, that the input  data cube \bartb{${\cal D}_{in}$} 
 has $k$ given measures $\mu_1,\mu_2,..., \mu_k$, and  that at some point in the OLAP
 process this cube is transformed to a cube \bartb{$\cal D$}, having measures 
 $\underline{\mu_1,\mu_2,..., \mu_k}; \tau_1,\tau_2, ...,\tau_l;\varphi, $
 where $\tau_1,\tau_2, ...,\tau_l$, with $l\geq 0$, are created measures and $\varphi$ is an input/output flag.

\subsection{Boolean Cell-selection Condition}\label{subsec:classical-prelim}
Before we start, we need to define the notion of a Boolean cell-selection condition, and  give a  lemma about its expressiveness 
we will use throughout Section~\ref{sec:classicalOLAP}.

  \begin{definition}[Boolean condition on cells]\rm\label{def:cell-selection}
Let $M({\cal D})=dom(D_1)\times dom(D_2)\times \cdots\times dom(D_d)$ be the matrix of $\cal D$.
A \emph{Boolean  condition on the cells of $M({\cal D})$} is a function $\phi$ from $M({\cal D})$ to $\{0,1\}$.
\bartb{We say that the cells of $M({\cal D})$ in the set $\phi^{-1}(\{1\})$ are \emph{selected} by $\phi$.}

We say that a Boolean condition $\phi$ is \emph{transformation-expressible} if there is a sequence of OLAP transformations 
$\tau_1,\tau_2,..., \tau_k$ such that $\phi(x_1,x_2,..., x_d)=\tau_k(x_1,x_2,..., x_d)$ for all $(x_1,x_2,..., x_d)\in M({\cal D})$.
\qedd
\end{definition}

\begin{lemma}\rm\label{lemma:boolean-closure}
If $\phi, \phi_1,\phi_2$ are transformation-expressible Boolean conditions on cells, then   
$\mbox{NOT } \phi$, $\phi_1\mbox{ AND } \phi_2$, and  $\phi_1\mbox{ OR } \phi_2$ are transformation-expressible Boolean conditions on cells. \qedd
\end{lemma}
   
\subsection{Dice}\label{subsec:Dice}

Intuitively, the \textit{Dice}
operation selects the cells in a cube $\cal D$ that satisfy a Boolean 
condition $\phi$ on the cells. The syntax for this operation is $\mbox{\sf DICE}({\cal D}, \phi),$ 
where  $\phi$  is a Boolean condition over level values
and measures. The resulting cube has the same dimensionality
as the original cube. This  operation is analogous 
to a selection in the relational algebra.
In  a data cube, it selects the cells that satisfy the 
condition $\phi$ by flagging  them with a $1$ in the output cube.
Our approach covers all typical cases 
in real-world OLAP~\cite{VZ14}. We next formalize the operator's definition in terms of our transformation language. In the remainder, 
we use the term  \textit{OLAP operation} to express a sequence of OLAP transformations.

 \begin{definition}[Dice]\rm \label{def:dice}
Given a data cube $\mathcal{D}$, the operation 
$\mbox{\sf DICE}({\cal D}, \phi),$ selects all cells of the matrix $M({\cal D})$ that satisfy the 
Boolean condition $\phi$ by giving them a $1$ flag in the output.
The condition $\phi$  
is a Boolean combination of conditions of the form: (a) 
  A selector on a value $b$ at a certain level $\ell$ of some dimension $D_i$;
 (b) A comparison condition  at some level $\ell$ from a dimension schema $\sigma(D_i)$ of a dimension $D_i$ of the cube of the form $\ell <c$ or $c<\ell $, where $c$ is a constant (at that level $\ell$); (c)
 An equality  or comparison condition on   some measure $\alpha$ of the form $\alpha=c$, $\alpha<c$ or $c<\alpha$, where $c$ is a (rational) constant.
\qedd
\end{definition}

\begin{property}\rm\label{prop:DICE}
Let $\cal D$ be a data cube en let $\phi$ be a Boolean condition on the cells of $M({\cal D})$ (as in Definition~\ref{def:dice}).
The \bartb{operation} $\mbox{\sf DICE}({\cal D}, \phi)$ is expressible \bartb{as an OLAP operation}. \qedd
\end{property}

\subsection{Slice}\label{subsec:Slice}
Intuitively, the  \textit{Slice}
operation    takes as input a $d$-dimensional, $k$-ary data cube $\cal D$ and a  dimension  $D_i$ and returns as 
 output $\mbox{\sf SLICE}({\cal D}, D_i)$,  
 which is a ``$(d-1)$-dimensional'' data cube in which the original measures $\mu_1,..., \mu_k$ are replaced by their
 aggregation (sum) over different values of elements in $dom(D_i)$. In other words, 
 dimension $D_i$ is removed from the data cube, and  will not be visible in the next operations. That means, 
 for instance, that we will not be able to dice on the levels of the removed dimension.
 As we will see, the ``removal'' of dimensions is, in our approach, implemented by   
 means of the destroyer measure $\delta$. 
 We remark that the aggregation above is due to the fact  
 that, in order to eliminate a dimension $D_i$, this dimension should have exactly one
 element~\cite{Agra97}, therefore a roll-up (which we explain later 
 in Section~\ref{subsec:RollUp}) to the level  
 \textit{All} in $D_i$ is performed.

\begin{definition}[Slice]\rm \label{def:SLICE}

Given a data cube $\mathcal{D}$,  and one of its dimensions $D_i$, 
the operation  $\mbox{\sf SLICE}({\cal D}, D_i)$ ``replaces'' the measures $\mu_1, \mu_2, ..., \mu_k$ by their aggregation (sum) 
${\mu_n}^{\Sigma_i}$ (for $1\leq n \leq k$) as:
${\mu_n}^{\Sigma_i} (x_1,...,x_{i-1},x_i, x_{i+1}, ..., x_d)=\sum_{x_i\in dom(D_i)} 
\mu_n(x_1,...,x_{i-1},x_i,x_{i+1}, ..., x_d),$  
for all $(x_1,...,x_{i-1},x_i, x_{i+1}, ..., x_d)\in M({\cal D})$.
 \bartb{Further, the operation $\mbox{\sf SLICE}({\cal D}, D_i)$ destroys all cells except those of the representative of $all$ for dimension $D_i$.}
We abbreviate the above $1$-di\-men\-sional sum as 
$ \mbox{\sc SUM}_{D_i}(\mu_n).$ \qedd
\end{definition}

 \begin{property}\rm\label{prop:SLICE}
Let $\mathcal{D}$ be a data cube  and let  $D_i$ be one of its dimensions.
The  \bartb{operation}  $\mbox{\sf SLICE}({\cal D}, D_i)$ is expressible as an OLAP operation. \qedd \end{property}

\begin{example}\rm\label{ex:SLICE}
Consider   dimensions $Product,$ $\ab Location,$ and $\ab Time$, and measure $\mu_1=sales,$ in our running example.  The operation  $\mbox{\sf SLICE}({\cal D}, Location)$ 
 returns   a cube with 
 $(product,time)$-cells  containing the sums of $\mu_1$ for each product-time combination,  over all location. All cells not belonging to the representative  of $all$ in the dimension $Location$ (i.e.,  $antwerp$), are destroyed. 
   The query is expressed by the following transformations.
 \begin{itemize}

\item $\tau_{l+1}=\gamma_{Product.Bottom}$ (prime labels on products);

 \item $\tau_{l+2}=\gamma_{Time.Bottom}$  (fresh prime labels on time moments);
  \item $\tau_{l+3}=\tau_{l+1}\cdot \tau_{l+2}$ (product of the two previous prime labels);

 \item $\tau_{l+4}=\mu_1\cdot\tau_{l+3}$ (product);
 \item $\tau_{l+5}=\mbox{\sc SUM}_{3}(\tau_{l+4})$ ($3$-dimensional sum);
 
 \item $\tau_{l+6}=\tau_{l+5}\mid_{\tau_{l+3}}$ (projection on prime product labels);
 \item $\tau_{l+7}=\sigma_{Location.All}$ (selects the representative of $all$ in the dimension $Location$);
 \item \bartb{$\delta=\tau_{l+7}$ (destroys all cells except  the representative of $all$ in  dimension $Location$);}
 \item $\varphi^{(1)} =\sigma_{Location.All}$ (this flag creation selects the relevant cells of the matrix).
\end{itemize}
 
Transformation $\tau_{l+4}$ gives each $(product,time)$-combination a unique prime product label. This label is multiplied by the $sales$
 in each cell. Then, $\tau_{l+5}$ is the global sum over $M({\cal D})$;   $\tau_{l+6}=\tau_{l+5}\mid_{\tau_{l+3}}$ 
 is the projection  over the prime product labels for $(product,time)$-combinations. This gives each cell above some fixed $(product,time)$-combination,
 the sum of the $sales$, over all locations, for that combination. All cells of  $M({\cal D})$  that do not belong to $antwerp$ (selected in  $\tau_{l+7}$), 
 which represents $all$, are destroyed by $\delta$.  
 \qedd
\end{example}



\subsection{Slice and Dice}\label{subsec:SandD}

A particular case of the \textit{Slice} operation occurs when the dimension to be removed already contains a unique value at the bottom level.
Then, we can avoid the roll-up to \textit{All}, and define a new operation, called \textit{Slice-and-Dice}. Although this can be seen as a \textit{Dice} operation 
followed by a $Slice$ one, in practice, both operations are
usually applied together. 

\begin{definition}\rm\label{def:SLICE-DICE}
\bartb{Given a data cube $\mathcal{D}$, one of its dimensions $D_i$ and some 
value $a$ in the domain $dom(D_i)$, 
the operation 
$\mbox{\sf SLICE-DICE}({\cal D}, D_i, a)$ contains  all the cells in the matrix $M({\cal D})$
 such that the value of the  dimension $D_i$ equals $a$. All other cells are destroyed. }
\qedd
\end{definition}

\begin{property}\rm\label{prop:SLICE-DICE}
Let $\cal D$ be a data cube, $D_i$ on of its dimensions  en let $a\in dom(D_i)$.
The operation $\mbox{\sf SLICE-DICE}({\cal D}, D_i, a)$ is expressible as an OLAP operation. \qedd 
\end{property}
 
\begin{example}\rm\label{ex:SLICE-DICE}
In our running example, the operation $\mbox{\sf SLICE-DICE}({\cal D}, Location, antwerp)$
is implemented by the output flag $\sigma_{Location.City=antwerp}$. 
\qedd
\end{example}

\subsection{Roll-Up and Drill-Down}\label{subsec:RollUp}
Intuitively, \textit{Roll-Up} aggregates measure values along a dimension up
to a certain level, whereas \textit{Drill-Down}  disagregates measure values   down to a   dimension level. Although at first sight it may appear that \textit{Drill-Down} is the inverse of  \textit{Roll-Up}~\cite{Agra97}, this is not always the case, e.g., if a \textit{Roll-Up} is followed by a   $Slice$ or a  $Dice$; here, we cannot just undo the \textit{Roll-Up}, but we need to account for the cells that have been eliminated on the way.

More precisely, the  \textit{Roll-Up} operation   takes as input a  data cube $\cal D$, a dimension $D_i$ and a subpath $h$ of a hierarchy $H$
over $D_i$, starting in a node $\ell'$ and ending in a node $\ell $, and returns  the aggregation of the original cube along  $D_i$ up to level $\ell$
for  some of the input measures $\alpha_1, \alpha_2, ..., \alpha_r$. \textit{Roll-Up} uses one of the classic SQL aggregation functions, applied to the indicated protected and computed measures
$\alpha_1, \alpha_2, ..., \alpha_r$ (selected from 
$\underline{\mu_1, \ab \mu_2, \ab ..., \ab \mu_k}; \tau_{1}, ..., \tau_{l}; \varphi$), namely  
 sum ($\mbox{\sf SUM}$),
 average ($\mbox{\sf AVG}$), 
 minimum /maximum ($\mbox{\sf MIN}$ and $\mbox{\sf MAX}$), 
 count and count-distinct ($\mbox{\sf COUNT}$ and $\mbox{\sf COUNT-DISTINCT}$). 
 Usually, measures have an associated \emph{default} aggregation function. The typical aggregation function for the measure $sales$, e.g., 
 is  $\mbox{\sf SUM}$.
We denote the above  operation as $\mbox{\sf ROLL-UP}({\cal D}, D_i, H(\ell'\rightarrow \ell), \{(\alpha_i,f_i)\mid i=1, 2, ..., r\}),$ 
where $f_i$ is one of the above  aggregation functions that is associated to $\alpha_i$, for $i=1,2,..., r$.
Since we are mainly interested in the expressiveness of this operation as a sequence of atomic transformations, only the destination node 
$\ell$ in the path $h$ is relevant. Indeed, the result of this roll-up remains the same if the subpath $h$ is extended to start from the $Bottom$ node 
of dimension $D_i$. So, we can simplify  the  notation, replacing $H(\ell'\rightarrow \ell)$ with 
$H(\ell),$  and assume that the roll-up starts at the  $Bottom$ level.

The  \textit{Drill-down} operation   takes as input a  data cube $\cal D$, a dimension $D_i$ and a subpath $h$ of a hierarchy $H$
over $D_i$, starting in a node $\ell$ and ending in a node $\ell' $ (at a lower 
level in the hierarchy), and returns  the aggregation of the original cube along  $D_i$ from the bottom level up to level $\ell'$.
The drill-down uses the same type of aggregation functions as the roll-up.
Again, since we are only interested in the expressiveness of this operation, the drill-down operation 
$\mbox{\sf DRILL-DOWN}({\cal D}, D_i, H(\ell'\leftarrow \ell), \{(\alpha_i,f_i)\mid i=1, 2, ..., r\}), $ 
has the same output as $\mbox{\sf ROLL-UP}({\cal D}, D_i, H(\ell'),  \{(\alpha_i,f_i)\mid i=1, 2, ..., r\}).$ 
\bartb{Therefore,   we can limit the further discussion in this section 
to the roll-up.}


\begin{definition}[Roll-Up]\rm\label{def:ROLLUP}
Given a data cube $\mathcal{D}$, one of its dimensions $D_i$, 
and a hierarchy $H$
over $D_i$, ending in a node $\ell $, 
the operation  $\mbox{\sf ROLL-UP}({\cal D}, D_i, H(\ell), \{(\alpha_i,f_i)\mid i=1,..., r\}) $ computes the aggregation of the measures $\alpha_i$  by their aggregation functions
$f_i$, for $i=1,2,..., r$, as follows:
 $$\displaylines{\quad {\alpha_i}^{f_i} (x_1,...,x_{i-1},x_i, x_{i+1}, ..., x_d)=\hfill{}\cr\hfill{}
 f_i(\{ \alpha_i ((x_1,...,x_{i-1},y_i, x_{i+1}, ..., x_d)\mid y_i\in dom(D_i) \mbox{ and } \rho_H(y_i,b) \} ),\quad}$$ 
for all $(x_1,...,x_{i-1},x_i, x_{i+1}, ..., x_d)\in M({\cal D})$, for which $ \rho_H(y_i,b)$, for some $b\in dom(D_i.\ell)$.
This roll-up flags all representative $Bottom$-level objects  as active.
\qedd
\end{definition}

\begin{property}\rm\label{prop:ROLLUP}
Let $\mathcal{D}$ be a data cube, let $D_i$ be one of its dimensions, 
and let  $H$ be a hierarchy over $D_i$ ending in a node $\ell$. Let $\{(\alpha_i,f_i)\mid i=1, 2, ..., r\}$ be a set of selected measures 
(taken from the protected measures $\mu_1, \ab \mu_2, \ab ..., \ab \mu_k$ and the computed measures $\tau_{1}, ..., \tau_{k}$
of $\mathcal{D}$), with their associated aggregation functions. 
The operation $\mbox{\sf ROLL-UP}({\cal D}, D_i, H(\ell), \{(\alpha_i,f_i)\mid i=1, 2, ..., r\}) $ is expressible as an OLAP operation. \qedd
\end{property}

\begin{example}\rm \label{ex:ROLLUP-1}
 We next express  the  \textit{Roll-Up} operation, using prime (product) labels, sums, projections, and    the $3$-dimensional sum. 
 We look at the query ``total sales per country''.
 We use the simplified syntax, only indicating the target level of the roll-up on the 
 \textit{Location} dimension (i.e., \textit{Country}). The query  $\mbox{\sf ROLL-UP}({\cal D}, Location ,Country, \{(sales, \mbox{\sf SUM)}\})$  is the result of the following transformations, given
 the measure $\mu_1=sales$:
 
 \begin{enumerate}
  \item $\tau_{\ell+1}=\gamma_{Product.Bottom}$ (prime labels on products);
 \item $\tau_{\ell+2}=\gamma_{Time.Bottom}$ (prime labels on time moments);
 \item $\tau_{\ell+3}=\gamma_{Location.Country}$ (prime labels on countries);
  \item $\tau_{\ell+4}=\tau_{\ell+1}\cdot \tau_{\ell+2}\cdot \tau_{\ell+3}$; (prime product label -- in one step);

 \item $\tau_{\ell+5}=\mu_1\cdot\tau_{\ell+4}$ (product of labels with $sales$);
 \item $\tau_{\ell+6}=\mbox{\sc SUM}_{3}(\tau_{\ell+5})$ ($3$-dimensional sum);
 
 \item $\tau_{\ell+7}=\tau_{\ell+5}\mid_{\tau_{\ell+4}}$ (projection on prime product labels);
  \item ${\varphi^{(1)}}=\sigma_{Location.Country}  $ (output flag on country-representatives).
  
\end{enumerate}
Transformation $\tau_{\ell+4}$ gives 
every product-date-country combination  a unique prime product label. Normally this product takes more steps. Above, we have abbreviated it to one transformation.
 The transformation  $\tau_{\ell+7}$ gives the aggregation result, and ${\varphi^{(1)}} $  is the flag that says that only the cities $antwerp$ and $paris$, 
 which represent the level $Country$, are active in the output
 (and nothing else of the original cube).   
 \qedd
\end{example}

\subsection{The Composition of Classical OLAP Operations}\label{subsec:compose-classical}
The main result of this paper is the proof of the completeness of an OLAP algebra, composed of the OLAP operations  Dice (Section~\ref{subsec:Dice}, 
Slice (Section~\ref{subsec:Slice}), Slice-and-Dice (Section~\ref{subsec:SandD}),
Roll-Up, and  Drill-Down (Section~\ref{subsec:RollUp}). This is summarized by Theorem \ref{theo:main}.

\begin{theorem}\rm \label{theo:main}
The classical OLAP operations and their composition are expressible by OLAP operations (that is, as sequences of atomic OLAP transformations).\qedd
\end{theorem}

We next  illustrate the power and generality of
our approach, combining a sequence of OLAP operations, and expressing them as a sequence of OLAP transformations.

\begin{example}\rm  \label{ex:ROLLUP-3}

An OLAP user  is analyzing sales in different countries and regions.
She wants to  compare sales in the north of Belgium (the Flanders region), and in the south of France (which we, generically, have denoted \textit{south} in our running example). She first filters the cube, keeping just the cells of those two   regions. This is done with the expression: $\mbox{\sf DICE}({\cal D}, Location.Region=flanders\  OR \ Location.Region=south).$ We showed that this can be implemented as a sequence of atomic OLAP transformations.
Now she has a cube with the  cells that  have not been destroyed. Next, within the same navigation process, she obtains the total sales  in France and Belgium,
only considering the desired regions, by means of:
$\mbox{\sf ROLL-UP}({\cal D}, Location ,Country, \{(sales, \mbox{\sf SUM)}\}).$ This will only consider the valid cells for rolling up. After this, our user only wants
to keep the sales in France. Thus, she writes:
$\mbox{\sf DICE}({\cal D}, Location.Country=france).$ Finally, she wants to go back to the details, one level below in the hierarchy, so she writes: $\mbox{\sf DRILL-DOWN}({\cal D}, Location ,Region, \{(sales, \mbox{\sf SUM)}\}),$  implemented as a roll-up from the bottom level to  \textit{Region},   only considering the cells that have not been destroyed. \qedd
\end{example}


\section{Conclusion and Discussion}\label{sec:Conclusion}

 We have presented a formal, mathematical approach, to solve a practical problem, which  
 is, to provide  a formal semantics to a collection of the OLAP  
 operations  most frequently used in real-world practice. Although OLAP is a very popular 
 field in data analytics, this is the first time a formalization like this is given.
 The need for this formalization is clear: in a world being flooded by data of  
 different  kinds, users must be  provided with tools allowing them to have an
 abstract ``cube view'' and cube manipulation capabilities, regardless of the 
  underlying data types. Without a solid basis and unambiguous definition of cube  
  operations, the former could not be achieved.  We claim that our work is the 
  first one of this kind, and will serve as a basis to build more robust 
  practical tools to address the forthcoming challenges in this field.  
   
  We have addressed the four core OLAP operations: slice, dice, roll-up, and drill-down.
  This does not harm the value of the work. On the contrary, this approach allows us to focus on our main  interest, that is, to study the formal basis of the problem. 
 Our line of work can be extended  to address other kinds of OLAP queries, like queries involving more complex aggregate functions like moving averages, rankings, and the like. Further, cube combination operations, like drill-across, must be included in the picture. We believe that our contribution  provides a solid basis upon which, a complete OLAP theory can be built.



 

{\vspace{.5cm} \ni{\bf Acknowledgements:}  
Alejandro Vaisman was  supported by a travel grant from Hasselt University (Korte verblijven--inkomende mobiliteit, BOF15KV13). He was also partially supported by 
PICT-2014 Project 0787.}  
\bibliographystyle{abbrv}

 \end{document}